\documentclass[11pt]{article}
\usepackage[letterpaper,margin=1in]{geometry}

\usepackage{times}
\usepackage{soul}
\usepackage{url}
\usepackage[hidelinks]{hyperref}
\usepackage[utf8]{inputenc}
\usepackage[small]{caption}
\usepackage{graphicx}
\usepackage{amsmath}
\usepackage{amsthm}
\usepackage{booktabs}
\usepackage{algorithm}
\usepackage{algorithmic}
\usepackage[switch]{lineno}

\usepackage{amssymb}
\usepackage{mathrsfs}
\usepackage{enumerate}
\usepackage{stackrel}
\usepackage{lipsum}
\usepackage[toc,page]{appendix}
\usepackage{natbib}
\usepackage{xcolor}
\newtheorem{definition}{Definition}
\newtheorem{prop}{Proposition}

\newtheorem{theorem}{Theorem}
\newtheorem{lemma}{Lemma}

\begin{document}




\begin{center}
{\LARGE\bf Improving Quantal Cognitive Hierarchy Model Through Iterative Population Learning}
\bigskip

{\large Yuhong Xu}\\
{\it School of Computing and Information Systems\\
Singapore Management University\\}
\texttt{yuhongxu@smu.edu.sg}
\medskip

{\large Shih-Fen Cheng\footnotemark
}\\
{\it School of Computing and Information Systems\\
Singapore Management University\\}
\texttt{sfcheng@smu.edu.sg}
\footnotetext{Corresponding Author: 80 Stamford Road, Singapore 178902. Phone: +65 6828-0526, Fax: +65 6828-0919.}
\medskip

{\large Xinyu Chen}\\
{\it School of Computing and Information Systems\\
Singapore Management University\\}
\texttt{xychen@smu.edu.sg}
\medskip

{\small \today}
\end{center}

\begin{abstract}
    In domains where agents interact strategically, game theory is applied widely to predict how agents would behave. However, game-theoretic predictions are based on the assumption that agents are fully rational and believe in equilibrium plays, which unfortunately are mostly not true when human decision makers are involved. To address this limitation, a number of behavioral game-theoretic models are defined to account for the limited rationality of human decision makers. The ``quantal cognitive hierarchy'' (QCH) model, which is one of the more recent models, is demonstrated to be the state-of-art model for predicting human behaviors in normal-form games. The QCH model assumes that agents in games can be both non-strategic (level-0) and strategic (level-$k$). For level-0 agents, they choose their strategies irrespective of other agents. For level-$k$ agents, they assume that other agents would be behaving at levels less than $k$ and best respond against them. However, an important assumption of the QCH model is that the distribution of agents' levels follows a Poisson distribution. In this paper, we relax this assumption and design a learning-based method at the population level to iteratively estimate the empirical distribution of agents' reasoning levels. By using a real-world dataset from the Swedish lowest unique positive integer game, we demonstrate how our refined QCH model and the iterative solution-seeking process can be used in providing a more accurate behavioral model for agents. This leads to better performance in fitting the real data and allows us to track an agent's progress in learning to play strategically over multiple rounds.
\end{abstract}

\section{Introduction}

When an agent's utility depends on not just its own action but also other agents' actions, it is well-known that these settings could be modeled and analyzed using game theory, and Nash equilibrium helps to predict how agents would behave collectively \citep{fudenberg1991game}. However, Nash equilibrium only works under the assumption that agents are fully ``rational'', as such, when human decision makers, who are at best partially rational, participate in a game, Nash equilibrium would provide poor outcome predictions \citep{goeree2001ten}. To address this issue, the field of behavioral game theory aims to explicitly model human agents' limited rationality, and come up with refined equilibrium concepts that would be more suitable for games with human agents.

One popular behavioral game-theoretic model is the ``cognitive hierarchy'' (CH) framework introduced by \citet{camerer2004cognitive}, which allows us to explicitly specify different rationality levels for agents in a game. In the CH framework, non-strategic agents are regarded as level-0, and their strategies are generated irrespective of other agents (e.g., uniformly randomly or greedily). For strategic agents at level-$k$ ($k \geq 1$), they would assume that other agents would be behaving at level-$j$, where $j < k$, and compute best responses\footnote{\citet{camerer2004cognitive} assume that an agent with level-$k$ believes that its opponent's reasoning levels would follow a Poisson distribution with parameter $\tau$ between 0 and $k-1$.}. 

In the literature, there are two major ways to improve the 
performance of the CH framework: 1) best response computation: instead of assuming that agents would choose best responses deterministically, assume that agents choose best responses stochastically according to each strategy's utility value; this idea is borrowed from \citet{mckelvey1995quantal}, and the resulting quantal cognitive hierarchy (QCH) model is demonstrated to match data involving human subjects better \citep{ostling2011testing,wright2017predicting}; 2) level-0 strategies: as strategic player's strategies are dependent on level-0 player's strategies, more realistic and sophisticated level-0 strategies could lead to better overall model performance, as demonstrated by \citet{wright2019level} and \citet{ho2021bayesian}.

In this paper, we look at the refinement of the opponent models for strategic agents as the third avenue for improving the CH framework. Our proposal is based on the observation that the computation of best responses for level-$k$ agents ($k > 1$) depends on the normalized belief distribution on the opponent's reasoning levels. By fitting an agent's decision traces to the level-$k$ best responses, we can then derive each agent's reasoning level
 distribution, which can be aggregated to the reasoning level distribution at the population level. Formally speaking, we are thus seeking a stable population reasoning level distribution which is the fixed point of the above iterative process. As our population reasoning level distribution is derived as a fixed point using the agent's actual behaviors, it should be more accurate than the commonly assumed Poisson distribution and should allow us to better measure the agent's true reasoning levels.

In studying the above process, we aim to make the following contributions:
\begin{itemize}
    \item We formally define the determination of the population reasoning level distribution as a fixed point-seeking problem.
    
    \item We demonstrate that a fixed point exists in the population dynamics that we define.
    
    \item We propose an iterative process that could efficiently identify a stable population reasoning level distribution.
    
    \item Finally, we test our approach on the lowest unique positive integer game dataset collected from a series of laboratory experiments. Against both the equilibrium and the state-of-the-art QCH model, we demonstrate that we can fit agents' behavioral traces much better (in Wasserstein distance, our approach outperforms the QCH model by close to 50\%). Our approach is also capable of producing estimations on individual agents' reasoning level distributions. Combined with personal-level observations, our approach can potentially be used as a way to accurately measure how strategically sophisticated are individual agents.
\end{itemize}

\section{Background: Behavioral Models}

As demonstrated by many researchers in behavioral game theory, the solution concept of Nash equilibrium from the classical game theory has not fared well when being tested in real-world settings, mainly due to human players' inability to perform strategic reasoning and their inaccurate belief about other agents (e.g, see arguments by \citet{camerer2004cognitive}). A wide variety of human behavioral models have thus been proposed to address the insufficiency of Nash equilibrium in predicting human behaviors. In this section, we introduce the most important behavioral models that appeared in the literature, with emphasis on the framework that we base our work on.

Based on how past researchers improve human behavioral models, we divide the literature into two categories: the ones that introduce explicit ``reasoning levels'', and the ones that introduce ``decision noises''.

\subsection{Reasoning Levels}

A key assumption behind Nash equilibrium is that players are capable of performing infinite iterations of strategic reasoning. Thus a straightforward way to relax this assumption is to explicitly incorporate the number of iterations humans can perform in the model. The first line of research that realizes this idea is the Level-$k$ (Lk) model \citep{stahl1994experimental,nagel1995unraveling,costa2001cognition}. The Cognitive Hierarchy (CH) model \citep{camerer2004cognitive}, which is more general, also follows the same idea. In both Lk and CH models, it is assumed that each player $i$ has an associated level of reasoning $k_i \in \{0,1,2,\ldots\}$, and a player with reasoning level $k$ can only best respond to players with lower levels. The major difference between Lk and CH models is that level-$k$ players in the Lk model only best respond to level-$(k-1)$ players, while level-$k$ players in the CH model best respond to all players with lower levels $(0, 1, \ldots, k-1)$. \citet{camerer2004cognitive} further suggest that the distribution of agent's levels should follow a Poisson distribution, and when a level-$k$ agent best responds to lower levels, she assumes that the distribution over all lower-level agents follows a truncated and normalized Poisson distribution up to $(k-1)$. We call such CH model the Poisson-CH model. 

\subsection{Decision Noises}

In standard games, we assume that players always follow best responses. To model noises in human decision making, a widely adopted idea is to introduce quantal responses, i.e., assuming agents would choose strategies stochastically in accordance with their payoff values: the higher/lower the payoff value, the higher/lower the probability of being chosen. \citet{mckelvey1995quantal} propose the quantal responses as a way to refine the equilibrium in games and demonstrate that the resulting quantal response equilibria fit better than Nash equilibria in many empirical games. By introducing quantal responses to the Lk model, the quntal Lk (QLK) model \citep{stahl1994experimental} is created, and is shown to provide a better fit than the deterministic Lk model. Similarly, the Poisson quantal CH (Poisson-QCH) model \citep{wright2012behavioral,wright2017predicting,wright2019level} is also created by introducing quantal responses to the Poisson-CH model. 
As our model is built based on the Poisson-QCH framework, we shall briefly introduce the notation and the fundamentals of the Poisson-QCH model.

\subsection{Quantal Cognitive Hierarchy}

Denote $G$, the game of interest, as a tuple $(N,A,u)$, where $N=\{1,\ldots,n\}$ is the finite set of agents, $A = A_1\times \ldots \times A_n$ is the set of joint action profile, with $A_i$ being agent $i$'s finite action set, and $u_i: A \rightarrow R$ is agent $i$'s utility function that maps a joint action profile to a real number. We define agent $i$'s mixed strategy space as $s_i = \Delta(A_i)$, which is the probability distribution over $A_i$. We define the joint strategy profile $s = s_1 \times \ldots \times s_n$, and we define $u_i(s)$ to be the expected utility of the mixed strategy $s$. We let $s_{-i}$ to be a strategy profile consisting of all agents except agent $i$. The mixed strategy profile $s^{*}$ is a Nash equilibrium if no player can improve her own expected payoff by deviating independently.

Following the definitions in \citet{wright2019level}, we define agent $i$'s \emph{quantal best response} to a strategy profile $s_{-i}$ as $QBR_i(s_{-i}; \lambda)$, which is a mixed strategy $s_i$ defined as:
\begin{align}
s_i(a_i) = \frac{\exp( \lambda \; u_i(a_i, s_{-i})) }{\sum_{a'_i} \exp( \lambda \; u_i(a'_i, s_{-i}) ) },\label{eqn:qbr}
\end{align}
where $\lambda$ is the \emph{precision} parameter indicating an agent's sensitivity to the actual utility value; agents choose uniformly if $\lambda \rightarrow 0$, and choose exact best responses if $\lambda \rightarrow \infty$.

Following \citet{camerer2004cognitive}'s Poisson assumption on agent's level distribution, and the notations by \citet{wright2019level}, we formally define the Poisson-QCH model as follows.
\begin{definition}[Poisson-QCH model, Definition 2 of \citet{wright2019level}]
    Let $\pi_{i,m} \in \Delta(A_i)$ be the mixed strategy predicted to be played by agent $i$ with level $m$. Let
    \begin{equation}
        \pi_{i,0:m} = \frac{\sum_{l=0}^{m} \text{Poisson}(l;\tau) \pi_{i,l}}{\sum_{l'=0}^{m} \text{Poisson}(l';\tau) } \label{eqn:poisson_m1}
    \end{equation}
    be the truncated distribution over actions predicted for an agent conditional on that agent having level $0 \leq l \leq m$. Let $\pi_{-i,0:m}$ be the truncated distribution over actions predicted for agents other than $i$, conditional on those agents having levels $0 \leq l \leq m$. With this, $\pi_{i,m}$ in the Poisson-QCH model is defined as:
    \begin{equation}
        \pi_{i,0}(a_i) = |A_i|^{-1}, 
        \pi_{i,m} = QBR_i(\pi_{-i,0:m-1}; \lambda).\label{eqn:poisson_m2}
    \end{equation}
    The overall predicted distribution of actions is thus a weighted sum over each level:
    \begin{equation}
        Pr(a_i | G, \tau, \lambda) = \sum_{l=0}^{\infty} \text{Poisson}(l;\tau) \pi_{i,l}(a_i). \quad \square
    \end{equation}

\end{definition}
With this, we are now ready to describe our approach. (For brevity, we refer to Poisson-QCH as QCH below.)

\section{Iterative Population Learning}

The QCH model defined by \citet{wright2019level} has been demonstrated to work well for a wide variety of datasets from the behavioral game theory literature. However, we do see an opportunity to further improve the model both in theory and practice. 

Our proposed approach is inspired by the observation that the computation of QBRs in \eqref{eqn:qbr} is closely dependent on the Poisson assumption on agent's reasoning levels, as illustrated in \eqref{eqn:poisson_m1} and \eqref{eqn:poisson_m2}. However, instead of assuming a fixed Poisson reasoning level distribution in \eqref{eqn:poisson_m1}, which might not describe the actual agent population well, we propose to fit agents' observed action frequencies to the computed QBRs, and estimate the empirical reasoning level distribution for each and every agent. If the newly updated reasoning level distribution is close enough to the previously used distribution, we would terminate the process; otherwise, we should repeat the above process by re-computing QBRs. 

As our approach \emph{iteratively} estimates reasoning level distribution while fitting agents' behavioral traces to the QCH model, we call our approach QCH-IPL, with IPL referring to the \emph{Iterative Population Learning}. Our proposed QCH-IPL approach has two major advantages:
\begin{enumerate}
    \item The population reasoning level distribution is decided solely by the observed agent behaviors.
    \item It estimates the reasoning level at the agent level, which allows us to track an agent's progression (learning) over time if data permits.
\end{enumerate}
We next formally define our approach.

\subsection{The QCH-IPL Approach}\label{sec:qch-ipl}

Let $L=\{0,1,\ldots,m\}$ be the set of reasoning levels to be considered. Let $p = (p_0, \ldots, p_m) = \Delta(L)$ be the reasoning level estimation at the population level, and $p_{l, i}$ be the probability that agent $i$ belonging to level $l$. Naturally, $\sum_{l \in L} p_{l,i} = 1, \forall i \in N$. 
The QCH-IPL approach is defined by the iterative steps below:
\begin{enumerate}
    \item \textbf{Initialization}: Let $t \leftarrow 1$ be the current iteration. Initialize $(p_0^t, \ldots, p_m^t)$ uniformly randomly.

    \item \textbf{Compute QBRs for all levels}: Following \eqref{eqn:poisson_m1} and \eqref{eqn:poisson_m2}, but with the simplifying assumption that the game is symmetric (agents have identical action space, and agent payoffs depend only on the strategy chosen and the other strategies employed).\footnote{1. Note that symmetry is not a requirement, just a simplification, and can be generalized easily. 2. In the behavioral game literature, we indeed observe that all studied games are symmetric.} We use $A_0$ to represent the common action space.
    \begin{eqnarray}
        \pi_{0:k}^{t+1} = \frac{\sum_{l=0}^{k} p_l^t \pi_{l}^{t+1}}{\sum_{l'=0}^{k} p_{l'}^t }, \forall k \in L \\
        \pi_{0}^{t+1}(a_0) = |A_0|^{-1}, \forall a_0 \in A_0 \\
        \pi_{k}^{t+1} = QBR\left(\times_{j=1}^{n-1} \pi_{0:k-1}^{t+1}; \lambda\right), \forall k=1,\ldots,m.
    \end{eqnarray}
    Note that $\times_{j=1}^{n-1} \pi_{0:k-1}^{t+1}$ is essentially $\pi_{-i,0:k-1}^{t+1}$ with symmetry assumption.

    \item \textbf{Fit each agent's behaviors to QBRs}:
    \begin{equation}
        \left(p_{0,i}^{t+1}, \ldots, p_{m,i}^{t+1}\right) = CLR\left( (\pi_{0}^{t+1}, \ldots, \pi_{m}^{t+1}), tr_i\right), \forall i \in N,
    \end{equation}
    where agent $i$'s behavioral traces are denoted as $tr_i$ and $CLR$ refers to \emph{constrained linear regression}, a function that estimates an agent's reasoning level distribution based on her behavioral traces. We will explain CLR shortly.

    \item \textbf{Aggregation}: We compute reasoning level estimation for the whole agent population by simple average:
    \begin{equation}
        p_l^{t+1} = \sum_{i \in N} p_{l,i}^{t+1} / n, \forall l \in L.
    \end{equation}

    \item \textbf{Checking for convergence}: If 
    $$
    \left\|(p_{0}^{t+1}, \ldots, p_{m}^{t+1}) - (p_{0}^{t}, \ldots, p_{m}^{t})\right\|_2 < \epsilon,
    $$
    terminate; otherwise, $t \leftarrow t+1$, repeat from 2.

\end{enumerate}


\subsection{Constrained Linear Regression}

In Step (3) of our QCH-IPL approach, we propose to use the constrained linear regression (CLR) approach to estimate agent $i$'s reasoning level using her behavioral traces. The CLR approach is appealing since it is based on regular linear regression, thus is intuitive, simple, and has low data requirement (this is important since most field experiments we see from the literature have only data points in 10s or 100s at the individual level). 

We define the response variable as agent $i$'s action frequency and the independent variables as QBRs at different levels. Intuitively speaking, we are trying to identify what would be the best convex combination of different QBRs, such that the resulting mixed distribution best resembles an agent's observed action frequency. We need additional constraints beyond regular linear regression since the weights represent probabilities thus need to sum up to 1 and stay non-negative.

Formally speaking, the CLR is solved using the following quadratic programming formulation:

\begin{align}
    \min & \sum_{a_0 \in A_0} \left( y_i(a_0) - \sum_{l \in L} \beta_{i,l} \pi_l (a_0) \right)^2\\
    \nonumber \text{s.t.}&\\
    \nonumber &\sum_{l \in L} \beta_{i,l} = 1, 
    \nonumber \beta_{i,l} \geq 0, \forall l \in L.
\end{align}

\subsection{Fixed Point of the QCH-IPL Process}

By combining all steps in Section \ref{sec:qch-ipl} functionally into a single composite function, we can examine our QCH-IPL approach from a mathematical perspective:
\begin{align}
    (\pi_0, \ldots, \pi_m) = F\Bigl( G\Bigl(\bigl( CLR\left( (\pi_0, \ldots, \pi_m), tr_i  \right) \bigr)_{i \in N}\Bigr) \Bigr), \label{eqn:fixed_pt}
\end{align}
where $G(\cdot)$ is the aggregation function in Step (4), and $F(\cdot)$ is the computation of QBR ($\pi_0$, \ldots, $\pi_m$) in Step (2). 

From above we can see that $(\pi_0, \ldots, \pi_m)$ is a fixed point of Equation \eqref{eqn:fixed_pt}. In other words, we are seeking a vector $(\pi_0, \ldots, \pi_m)$, which would be returned as output when given as input. Below we provide proof that such a fixed point exists. 

\begin{theorem}
A fixed point exists in \eqref{eqn:fixed_pt}. \label{thm:fixed_point}
\end{theorem}
\begin{flushleft}
(Proof Sketch)
\end{flushleft}
Similar to \citet{kaufman1998user}, our existence proof is based on the \emph{Schauder-Tychonoff fixed point theorem}, and we follow the three steps below: 
\begin{itemize}
    \item \textbf{Step 1:} We represent the set of strategies in topology form. 
    \item \textbf{Step 2:} We establish the compactness and convexity properties of the strategy set. 
    \item \textbf{Step 3:} We establish the continuity property of the combined strategy/reasoning level iteration $F(G(\cdot,tr_i))$. 
\end{itemize}
The Schauder-Tychonoff fixed point theorem can be applied after these three steps are proved. 
These three steps are proved as lemmas in Appendix A. \quad $\square$

\bigskip
The existence of fixed point might not be applicable in the sense that the topology is not metrizable, since it relies on giving $\Pi$ the Cartesian product topology over its uncountable indices, such as uncountably many time points \citep{kaufman1998user}.  Contraction property needs to be established for the strategy generation $F$ to make sure the successive approximation executed by the proposed loop algorithm converges to a fixed point from any starting point, and this fixed point is unique \citep{kaufman1998user,saaty1964nonlinear}. To establish the property of contraction, we impose three additional restrictions on the strategy set: 
\begin{itemize}
    \item \textbf{Restriction 1:} Only strategies that vary continuously over time are considered. 
    \item \textbf{Restriction 2:} Only strategies that are equicontinuous are considered. 
    \item \textbf{Restriction 3:} The strategy set is closed. 
\end{itemize}
Denote the new strategy set after imposing the above three restrictions as $\Pi^*$, we can prove the existence of fixed points in $\Pi^*$ of the combined strategy/reasoning level iteration $F(G(\cdot,tr_N))$. More details can be found in Appendix B.

\section{Numerical Experiments}


Testing behavioral game theory models in the field is challenging in general. It is even more so for our QCH-IPL approach, as we need observations of individual agents' behavioral traces. The study by \citet{ostling2011testing} is one rare such instance where both large-scale field data and corresponding laboratory experiments at the individual level exist. We thus utilize the extensive laboratory data they have collected and openly shared for our numerical experiments.

For the remainder of this section, we first introduce the Swedish lowest unique positive integer (LUPI) game studied by \citet{ostling2011testing}, its connection to the Poisson game \citep{myerson1998population,myerson2000large} in the literature, and the Poisson-Nash equilibrium predicted by the theory. We then summarize the study performed by \citet{ostling2011testing}, which compares Poisson-Nash equilibrium against the CH model. We compare the performance of our QCH-IPL approach against all the baseline methods. We conclude by highlighting the additional insights we obtained, and the discussion on the strength and the weakness of our approach.

\subsection{The LUPI Game}

Following the assumptions by \citet{ostling2011testing}, the lowest unique positive integer (LUPI) game is defined as a Poisson game \citep{myerson1998population,myerson2000large} where the number of agents follows a Poisson distribution with a known average. All agents in the LUPI game choose integers from 1 to $K$ simultaneously, with the agent who chooses the lowest unique number winning. In the theoretical analysis, agents are assumed to be fully rational, best responding, and with equilibrium beliefs. As argued by \citet{ostling2011testing} and \citet{myerson1998population}, the Poisson assumption is crucial, as it greatly simplifies theoretical analysis. We should not repeat the whole analysis by \citet{ostling2011testing}, but we summarize the most important theoretical results below. 

\begin{prop}[Proposition 1 of \citet{ostling2011testing}]
There is a unique mixed equilibrium $(p_1, p_2, \ldots, p_K)$ in the LUPI game that satisfies the following properties: 
\begin{enumerate}
    \item Full support: $p_k>0$ for all $k$; 
    \item Decreasing probabilities: $p_{k+1} < p_k$ for all $k$;
    \item Convexity/concavity: $(p_{k+1} - p_{k+2}) > (p_{k} - p_{k+1})$ for $p_{k+1} > 1/n$, and $(p_{k+1} - p_{k+2}) < (p_{k} - p_{k+1})$ for $1/n > p_{k}$.
\end{enumerate}
The remaining less important properties are omitted.\label{prop:p_nash}
\end{prop}

\subsubsection{The Field and the Lab LUPI Games}

The field version of the LUPI game was launched in Sweden by the government-owned gambling monopoly Svenska Spel on January 29, 2007. In the field LUPI game, players pay SEK 10 (roughly 1 Euro) to participate and choose an integer between 1 and 99,999. The winning number, which is determined and announced daily, is the smallest and least frequently chosen number. Players choosing the winning number share the first prize. There are also second and third prizes offered. In total the field LUPI game run for 49 days (7 weeks).

The field LUPI game is different from the theoretical LUPI game in three aspects: firstly, instead of choosing the smallest unique number, players who choose the smallest and least-frequently number share the prize; secondly, players are allowed to choose up to six numbers; finally, there are second and third prizes. 

Recognizing these departures, \citet{ostling2011testing} decide to design a lab LUPI game that resembles the field LUPI game, yet restores the most important features of the theoretical LUPI game, namely: 1) the number of players follows Poisson distribution, 2) only unique least number wins the game; in the case where there is no unique least number, no one wins, and 3) players are only allowed to choose one number on their own (in the field LUPI game, players can utilize random number generator to generate their entries). To match the length of the field LUPI game, participants are asked to play 49 rounds. For each round, the winner (if any) earns \$7. Considering that the number of participants in the lab LUPI game is significantly fewer, the number range is reduced to between 1 and 99.
The winning number is made known to all participants after each round concludes. The winner per round is notified privately. There are in total 38 participants in the lab LUPI game, and the number of active players is drawn from a Poisson distribution with a mean of 26.9 in each round. 
Players not chosen to be active are still required to submit numbers. 

Participants are graduate and undergraduate students recruited at the University of California, Los Angeles, mostly aged between 18 to 22, with roughly equal males and females. Participants have various levels of exposure to game theory, half of them never participated in a lottery game before and few had heard of a similar game before this experiment. All sessions lasted for less than one hour and subjects earned a show-up fee of \$8 or \$13 in addition to earnings from the experiment (mean earnings equals \$8.60). To compare with the model performance in \citep{ostling2011testing}, we adopt their reference, where they describe experimental rounds as \emph{days} and seven-round intervals as \emph{weeks}. 

The complete dataset and the code can be downloaded from \url{https://www.aeaweb.org/articles?id=10.1257/mic.3.3.1}.

\subsection{Baselines}

When evaluating the performances of different behavioral models, \citet{ostling2011testing} study two baselines: the Poisson-Nash equilibrium and the QCH model (although \citet{ostling2011testing} do not explicitly call it a QCH model, they have implemented the \emph{CH model with quantal responses}, which is essentially the QCH model). Before delving into the actual performance analysis, we describe how these two baselines are computed.

\subsubsection{Poisson-Nash Equilibrium (PNE)}
The PNE can be computed by utilizing the following equilibrium conditions \citep{ostling2011testing}:
\begin{enumerate}
    \item If both $k$ and $k+1$ are chosen with positive probability in equilibrium, then: 
    $$p_k - p_{k+1} = - \frac{1}{n} \ln \left( 1 - n \cdot p_k \cdot \exp(-n \cdot p_k) \right).$$
    
    \item $\sum_{k=1}^{K} p_k = 1$.
    
    \item The expected payoff for playing numbers where $p_k=0$ should not be higher than those with positive probability. 
\end{enumerate}

\subsubsection{CH Model with Quantal Response (the QCH Model)}
There are two parameters in the QCH model: 1) $\tau$: the average reasoning level, assuming that the level distribution follows the Poisson distribution, and 2) $\lambda$: the precision parameter in the quantal response model. Due to insufficient data in the lab setting, \citet{ostling2011testing} choose to fix $\tau$ to the value estimated from the field data and search for $\lambda$ using maximum likelihood estimation (MLE) with a very fine grid search (they cut the search space into 100,000 grids). The QCH model, as demonstrated and argued by \citet{wright2017predicting}, is considered the current state-of-the-art model for fitting and predicting human behaviors in normal-form games. Although the game studied by \citet{ostling2011testing} is a Poisson game, the QCH model still fits the behavioral data well.

\subsection{Solving Lab LUPI Games with QCH-IPL}

Compared to the original QCH model, our QCH-IPL eliminates the need for $\tau$, since we iteratively learn the reasoning level distribution using empirical data. However, as our model still has the quantal component, we need to estimate $\lambda$ with MLE. We follow the approach by \citet{ostling2011testing} and search for $\lambda$'s value in the range of 1 to 20, in which we divide the search space into up to 500 grids. Our current results can be further improved if we adopt finer grids.
We execute our QCH-IPL for up to 100 iterations. If the convergence is not achieved after 100 iterations, we return the distribution with the best log-likelihood value.

\subsection{Performance Metrics}

\subsubsection{Log-likelihood}
Log-likelihood function is a logarithmic transformation of the likelihood function, which is the joint probability of the observed data viewed as a function of the parameters of the chosen statistical model.
The log-likelihood value is a classical measure of the goodness of fit of a model (how well a given model fits the observations): larger values correspond to better fit. While the actual log-likelihood value for a given model is mostly meaningless, it is useful for comparing two or more models fit over the same dataset.


\subsubsection{Chi-squared Goodness of Fit Test}
A Chi-squared test is a statistical hypothesis test for assessing the goodness of fit of the model. It establishes whether an observed frequency distribution differs from a theoretical distribution. The two hypotheses to be evaluated are: 1) the null hypothesis ($H_0$): the population follows the specified distribution; 2) the alternative hypothesis ($H_a$): the population does not follow the specified distribution.

The value of the test statistic is defined as:
\begin{equation}
    \chi^2=\sum_{i=1}^{n}{\frac{(O_i-E_i)^2}{E_i}},
    \label{eqn:chi}
\end{equation}
where $\chi^2$ is Pearson's cumulative test statistics that asymptotically approaches a $\chi^2$ distribution, $O_i$ is the number of observations of type $i$, and $E_i$ is the expected count of type $i$. The test statistic is compared against a critical value to decide whether the null hypothesis can be rejected. If the $\chi^2$ value is greater than the critical value, the $H_0$ is rejected, indicating the model does not provide a good fit for the data.

To compare our results with the ones from \citet{ostling2011testing}, we set the bin size to be 2 (i.e., 2 numbers per category), and execute the Chi-squared test using the average frequencies for each category. There are only 6 bins (up to number 12) included in the Chi-squared test to prevent the number of observations to drop below 5. Increasing the number of bins does not change the results of the test.

\begin{table*}[htb]
\caption{Goodness-of-fit for the laboratory data: Comparing Poisson-Nash equilibrium, QCH model, and QCH-IPL model.
}
\label{tab:main_results}
\centering
\setlength{\tabcolsep}{10pt}
\resizebox{\textwidth}{!}{%
\begin{tabular}{lcccccccc}
\hline
Week & (1) & (2) & (3) & (4) & (5) & (6) & (7) & Average \\
\hline
\textbf{Poisson-Nash equilibrium}   & & & & & & & & \\ 
$\chi^2$ (for average frequency) & 24.69 & 24.14 & 18.86 & 21.81 & 20.17 & 11.58 & 21.39 & -\\
(Degree of freedom)               & (5)***& (5)***& (5)***& (5)***& (5)***& (5)** & (5)***& - \\
Proportion below (percent)        & 82.25 & 88.55 & 87.61 & 88.64 & 88.64 & 92.86 & 87.06 & 87.94\\
Wasserstein distance                & 3.3624  &  1.1573  &  1.1004  &  1.3227   &  1.0377   &  1.0248  &  0.8960  & 1.4145\\
\textbf{QCH Model}  & & & & & & & & \\ 
Log-likelihood                    & –210.4 & –104.3 & –88.6 & –88.7 & –87.5 & –80.2 & –99.4 & -\\
$\tau$ (from field)               & 1.80 & 3.17 & 4.17 & 4.64 & 5.02 & 6.76 & 6.12 & -\\
$\lambda$                          & 1.26 & 5.97 & 16.89 & 5.59 & 5.28 & 22.69 & 4.52 & -\\
$\chi^2$ (for average frequency) & 24.31 & 18.80 & 8.49 & 4.57 & 6.83 & 2.74 & 10.06 & -\\
(Degree of freedom)               & (5)***& (5)***& (5) & (5) & (5) & (5) & (5)* & -\\
Proportion below (percent)        & 84.62 & 87.44 & 90.52 & 92.54 & 92.42 & 91.11 & 91.07 & 89.96\\ 

Wasserstein distance                & 4.9716   &  1.8493   & 0.2904 &  0.6884  &  0.5954  &  0.3234 & 0.5903  & 1.3298\\
\multicolumn{8}{l}{\textbf{QCH-IPL Model}}\\
Log-likelihood                    &  -185.0 & -67.0 & -72.9 & -78.2 & -78.9 & -69.7 & -85.3 & -\\
$\lambda$                         & 1.2 & 17.9 & 12.0 & 15.4 & 12.6 & 16.8 & 9.0  & - \\
$\chi^2$ (for average frequency) & 16.65 & 0.22 & 1.09 & 0.60 & 0.33 & 0.66  & 3.26 & -\\
(Degree of freedom)               & (5)*** & (5)  & (5)  & (5)  & (5)  & (5)  & (5) & -\\
Proportion below (percent)        & 86.58 & 96.07 & 95.74 & 95.19 & 95.10 & 96.41 & 93.70 & 94.11\\
Wasserstein distance                &  2.8693  &  0.3713  &  0.2149  &  0.4315  &  0.3412  &  0.3083 &  0.4353  & 0.7103\\
\hline
\end{tabular}}
\caption*{Note: The degree of freedom for a $\chi^2$ test is the number of bins minus one. The ``proportion below'' refers to the fraction of the empirical density that lies below the theoretical prediction. (Significance level: *** 1\%. ** 5\%. * 10\%.)}
\end{table*}

\begin{figure*}[htb!]
  \centering
    \includegraphics[width=.24\textwidth]{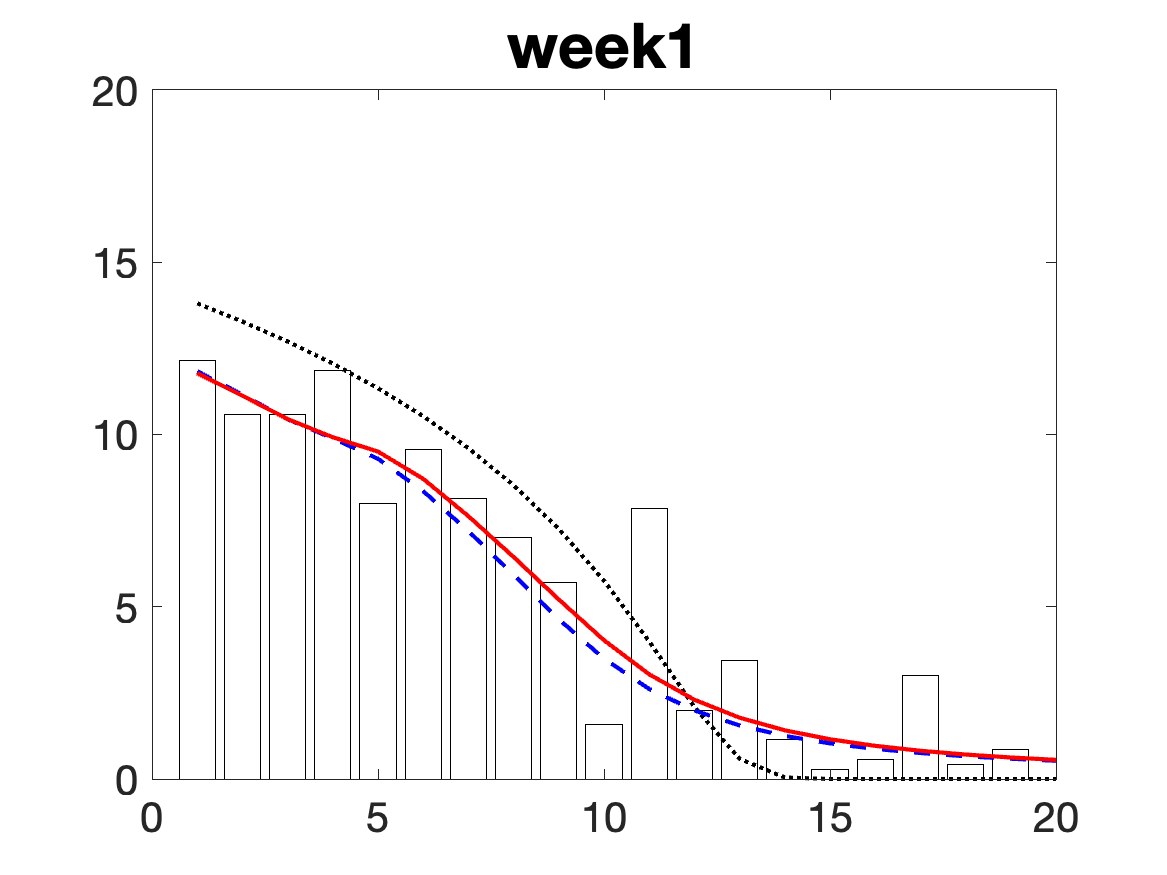}
    \includegraphics[width=.24\textwidth]{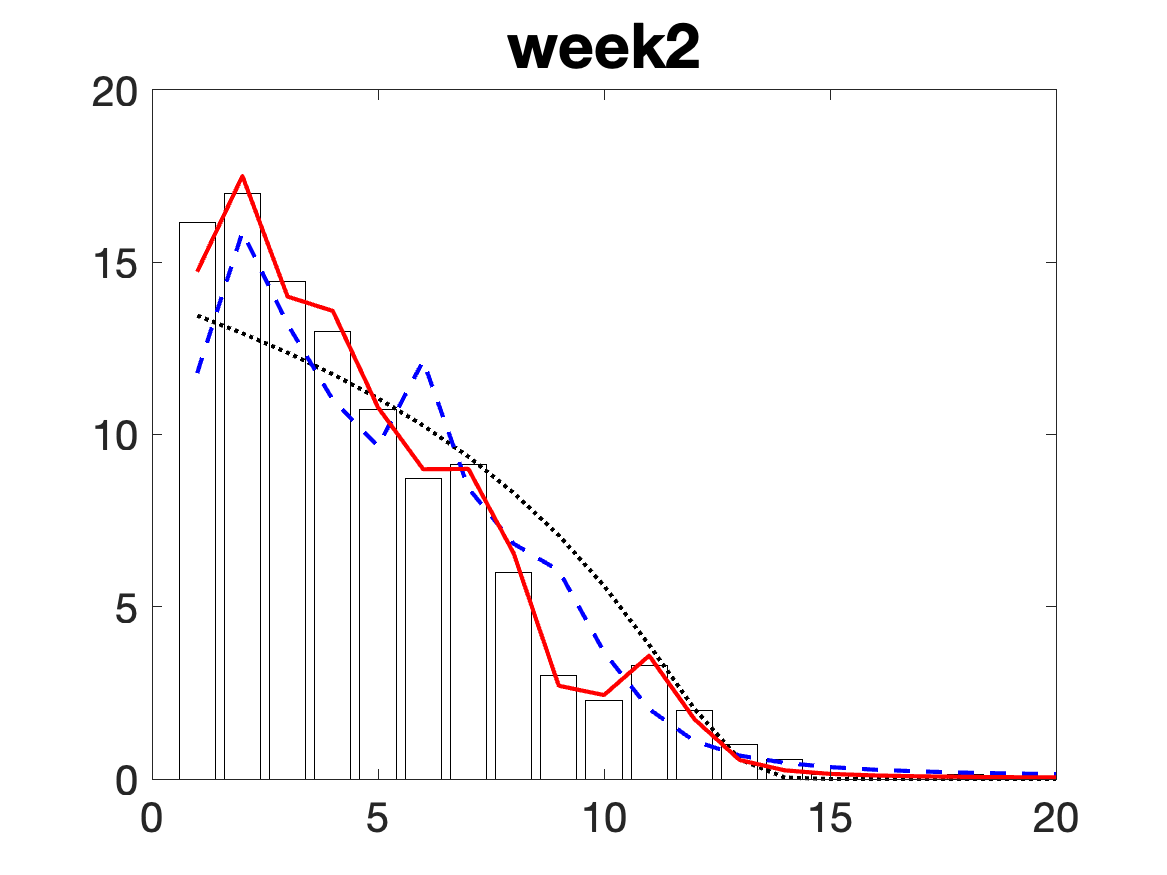}
    \includegraphics[width=.24\textwidth]{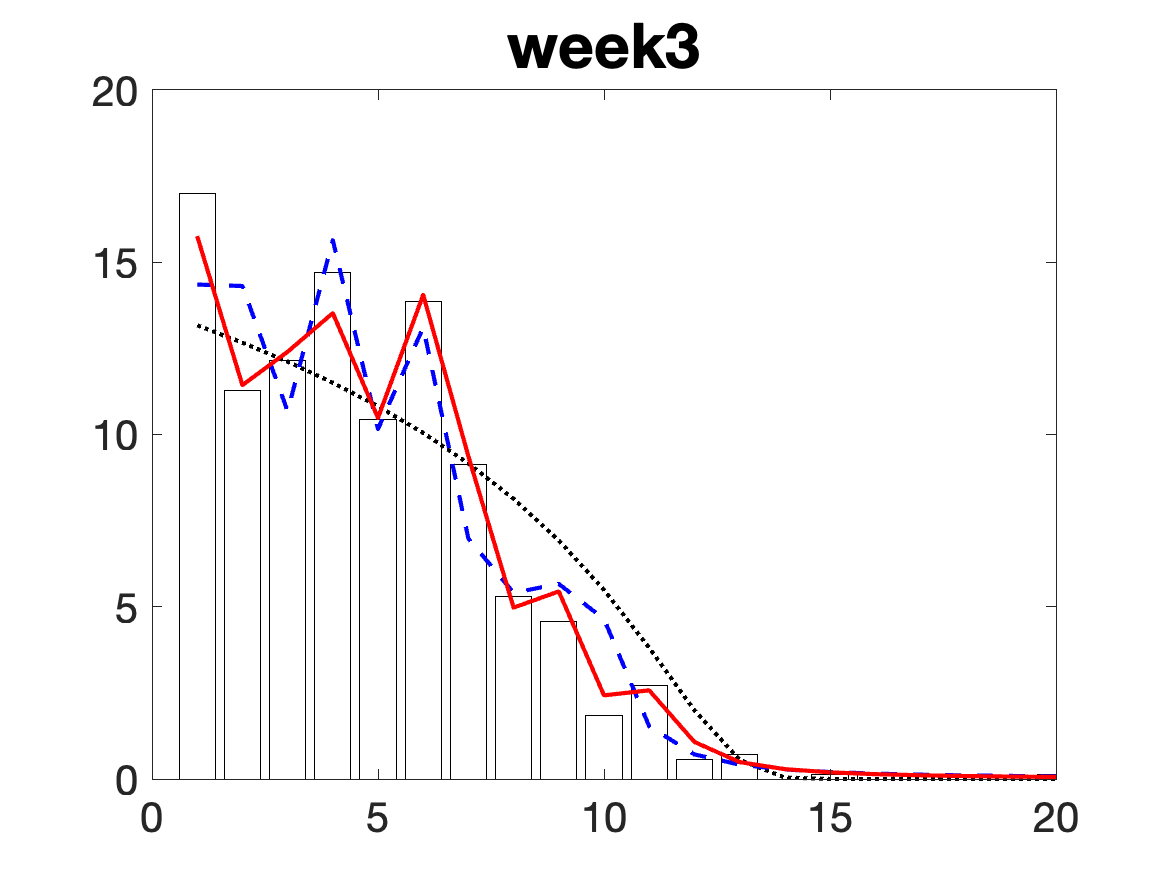}
    \includegraphics[width=.24\textwidth]{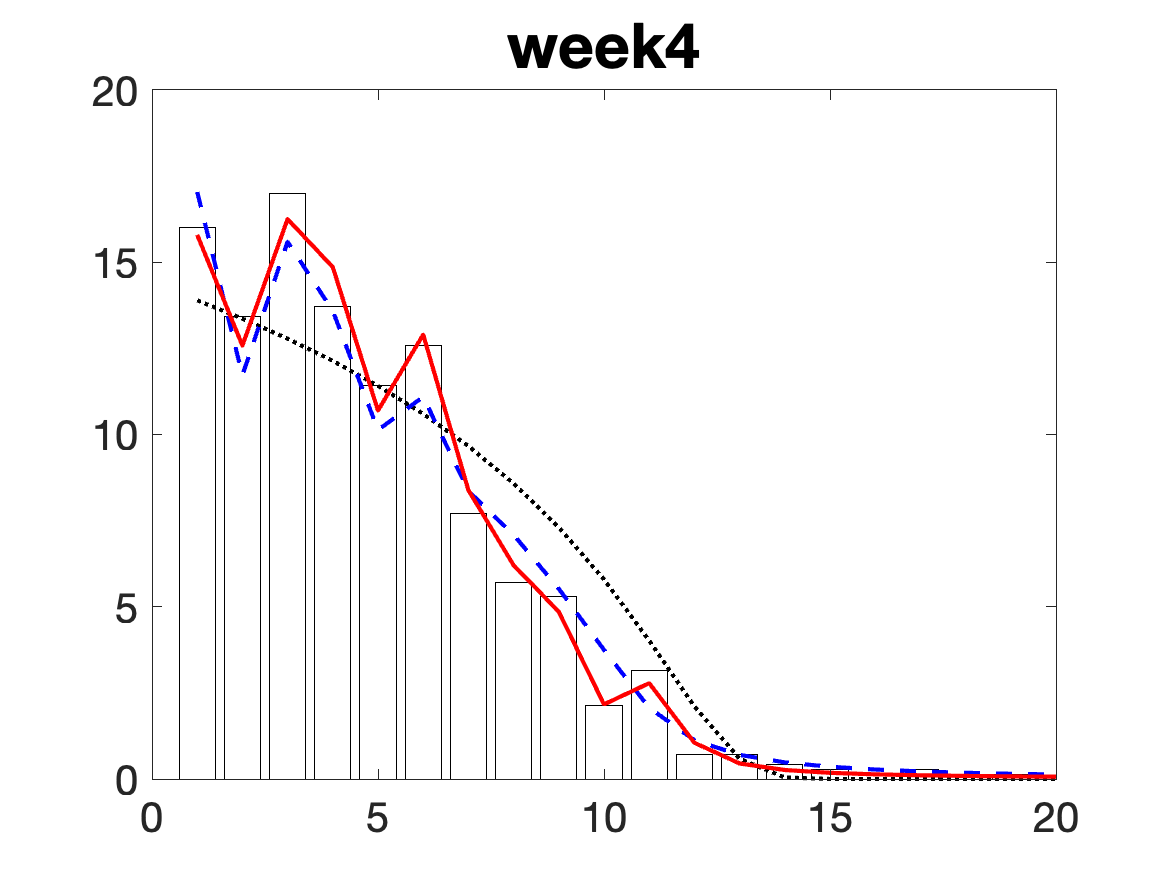}\\
    \includegraphics[width=.24\textwidth]{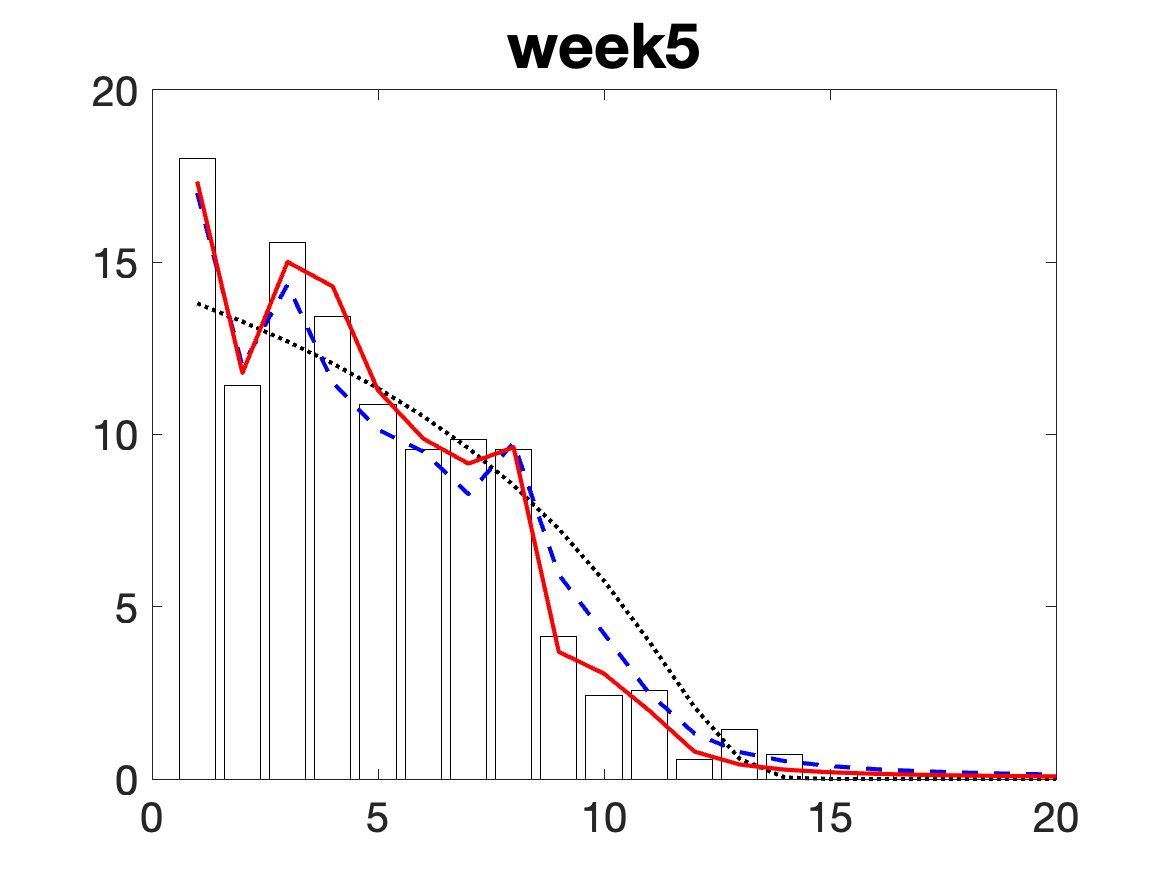}
    \includegraphics[width=.24\textwidth]{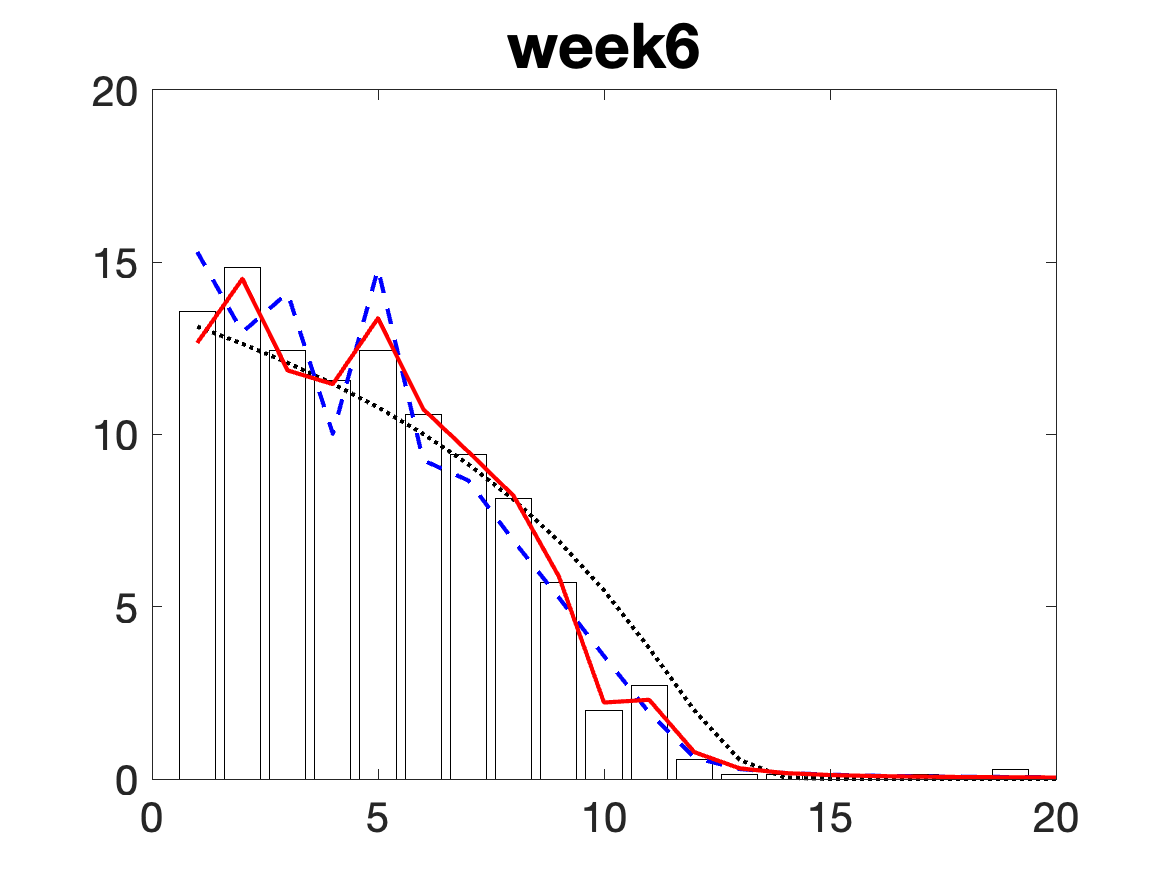}
    \includegraphics[width=.24\textwidth]{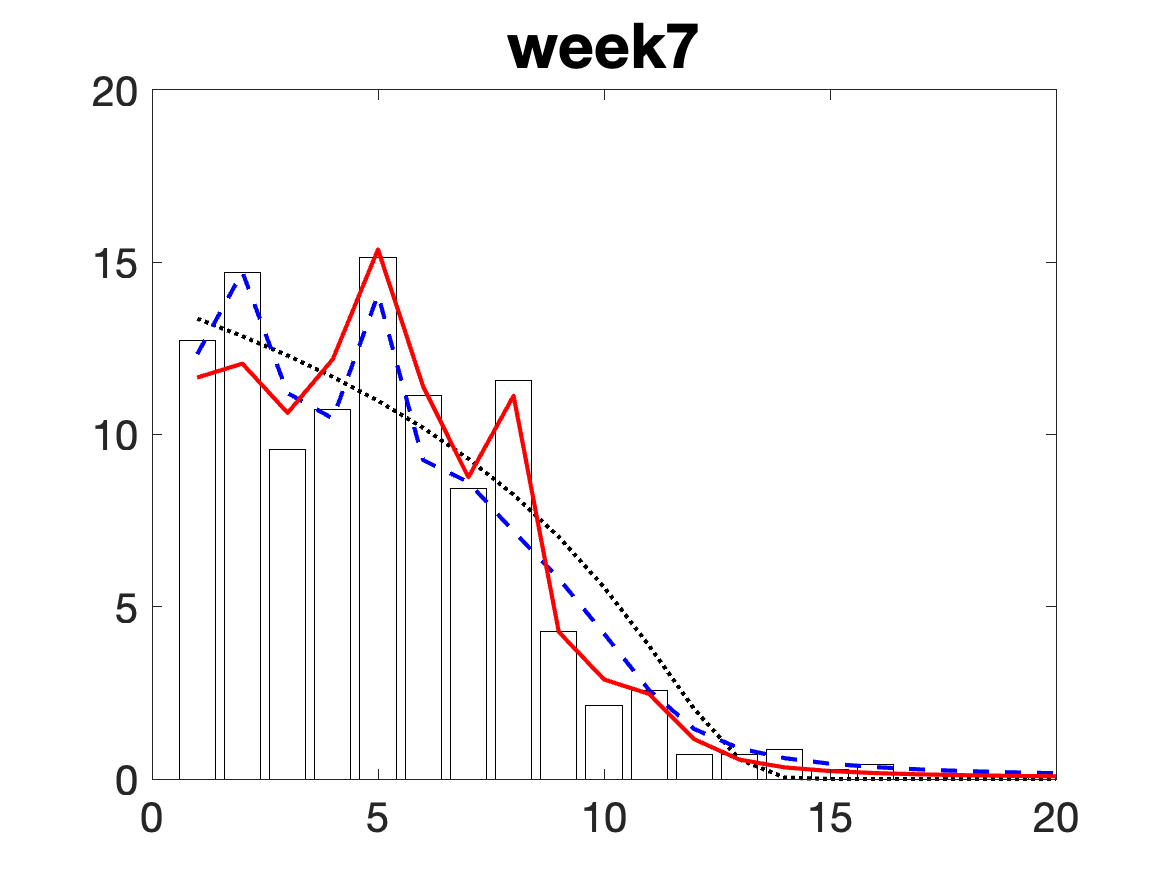}
    \includegraphics[width=.24\textwidth]{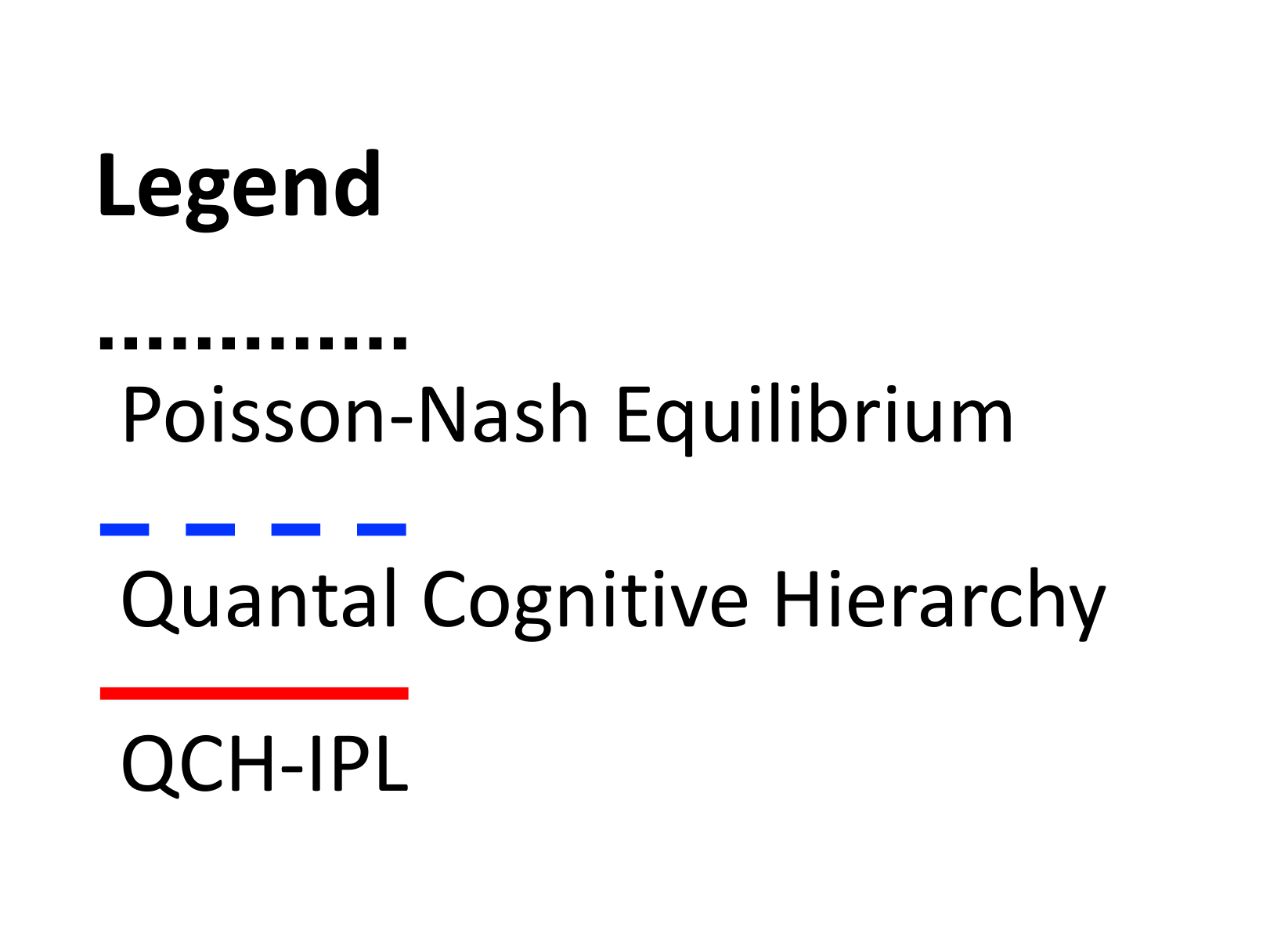}
    \caption{ Average daily frequencies of numbers chosen by players in the lab for each week, together with Poisson-Nash equilibrium prediction, and estimated QCH and QCH-IPL models.}
    \label{fig:weekly_hist}
\end{figure*}

\subsubsection{Wasserstein Distance}

Wasserstein distance (W-distance), more commonly known as the earth mover's distance (EMD) in computer science, is a distance measure between probability distributions. Intuitively speaking, the W-distance measures the effort to transform a probability distribution into another probability distribution. As the W-distance more accurately describes how two distributions are different, it has been increasingly used in computer science and particularly generative machine learning models \citep{arjovsky2017wasserstein}. In our main results, We utilize the W-distance to measure the difference between players' observed action frequencies and the distributions predicted by competing models.

\subsection{Main Results}

In this section, we present our main results together with the findings in \citet{ostling2011testing}, where they provide the goodness-of-fit of the PNE and the QCH model based on the laboratory data. The main results are summarized in Table~\ref{tab:main_results}. 

We first examine the Chi-squared test and the log-likelihood values, which are widely accepted measures of goodness of fit in econometrics. In all weeks of the lab experiments, the PNE is rejected with high significance (all at 1\% level, except week 6, which is at 5\% level; this strongly suggests that the PNE is not a good fit for the data). For the QCH model, only weeks 1 and 2 are rejected at 1\% level and week 7 at 10\% level. For our QCH-IPL model, only week 1 is rejected at 5\% level. From this, we learn that although players' actual action choice frequencies may look not too different from the PNE (the ``proportion below'' measure seems to be reasonably high, above 80\% in all weeks), player strategies cannot be sufficiently described by the PNE statistically. By this measure, the QCH model is much more representative statistically than the PNE, and the QCH-IPL model is even better. To compare the goodness of fit for the QCH and QCH-IPL models, we further look at the log-likelihood values. In all weeks, the QCH-IPL model performs better than the QCH model by a large margin.

The ``proportion below in percentage'' is another measure used by \citet{ostling2011testing} to estimate how much probability masses of players' action frequencies are captured by the different behavioral models. By this measure, the QCH-IPL model outperforms both PNE and the QCH model over all weeks, and on average the advantage margins are 7\% and 4.6\% respectively.

Finally, to more accurately quantify the difference between players' action frequencies and the distributions estimated by behavioral models, we utilize the W-distance as a new performance measure. We again see that on average the QCH-IPL model outperforms both the PNE and the QCH model, by big margins of 49.8\% and 46.6\% respectively.

To visualize the behavioral models and players' actual action frequencies, we provide plots for all weeks in Figure~\ref{fig:weekly_hist} (as numbers above 20 have low usage frequencies, we cut the x-axis off at 20, so that we can focus on the significant portion of the histogram). In each week's subplot, we use bars to represent players' choice frequencies for numbers from 1 to 20. On top of the histogram, we also plot distributions estimated by the PNE (as dotted lines), QCH model (as dash lines), and QCH-IPL model (as solid lines). From Figure~\ref{fig:weekly_hist} we can see that the PNE is mostly smooth, thus not able to capture discontinuities in players' action frequencies. Both QCH and QCH-IPL models respond to discontinuities, but QCH-IPL avoids most overshoots or undershoots made by the QCH model.

\subsection{Behavioral Analysis at the Individual Level}

One of the strengths of our QCH-IPL approach is that we can estimate the reasoning level distribution at the individual level. This allows us to derive personalized behavioral models, from which we can not only quantitatively capture diverse subject behaviors but also track the evolution of subjects' behavioral sophistication over time. To demonstrate this we provide average reasoning level estimations for all lab experiment participants over all weeks in Appendix C.


\section{Related Work}

Human behaviors have been studied widely in many domains. A major focus has been the effect of cognitive ability and various factors that influence one's cognitive ability \citep{falch2011effect,rushton2005thirty,burnham2009higher,tobias2003effects,o2001forging}. Another stream of thought roots in the game theory literature, and looks for ways to improve the mathematically rigorous game theoretic models, so that they fit actual human behaviors better. For example, the quantal response equilibrium (QRE) \citep{mckelvey1995quantal}, noisy introspection model (NI) \citep{goeree2001ten}, the cognitive hierarchy model (CH) \citep{camerer2004cognitive}, the closely-related level-$k$ (Lk) model \citep{costa2001cognition,nagel1995unraveling}, quantal level-$k$ (QLk) model \citep{stahl1994experimental}, and the quantal cognitive hierarchy (QCH) model \citep{ostling2011testing,wright2017predicting}. \citet{wright2017predicting} have recently studied and compared a wide variety of the LK and CH variants. This line of research is most relevant to our paper.

\section{Conclusions}

In this paper, we propose the QCH-IPL model, which is based on the state-of-the-art QCH model from the behavioral game theory literature. We relax the assumption that agents' reasoning levels follow the Poisson distribution, and propose to exploit the mutual dependency between the reasoning level distribution and the best response computation. This dependency exists since the computation of best responses depends on the reasoning level distribution, yet with computed best responses for all levels, we can fit them to an agent's observed behavioral traces to estimate this agent's reasoning level distribution. We test our QCH-IPL model in the Swedish lowest unique positive integer game conducted in a lab setting. We demonstrate how our QCH-IPL approach can be used in providing a more accurate behavioral model for agents. By computing the Wasserstein distances between the empirical choice distributions and the choice frequencies predicted by competing models, we see that our QCH-IPL approach outperforms both PNE and QCH models by close to 50\%. Besides better performance in fitting agents' behaviors, our QCH-IPL approach also allows us to track individual agents' progress in learning to play strategically over multiple rounds.

\section*{Acknowledgement}
This research is supported by the Ministry of Education, Singapore, under its Social Science Research Thematic Grant (Grant Number MOE2020-SSRTG-018). This work was presented at various seminars and we would like to thank participants for their comments. A special thanks goes to Prof. Joseph Tao-yi Wang for pointing us towards the LUPI dataset and suggesting ways we could improve the paper. 

\bibliographystyle{plainnat}
\bibliography{rh_bib}

\newpage
\appendix

\section{Existence Proof of Theorem \ref{thm:fixed_point}}\label{A}

\subsection{Fixed Point Theorems}
We first state three well-known fixed point theorems that are utilized in our proof.

\begin{lemma}
\label{lem_1}
(Schauder-Tychonoff) Let $C$ be a compact convex subset of a locally convex linear topological space. If $f: C \rightarrow C$ is continuous, then there exists a fixed point of $f$ in $C$.
\end{lemma}

\begin{lemma}
\label{lem_2}
(Schauder) Let $Y$ be a compact convex subset of a complete metric space. If $f: Y \rightarrow Y$ is continuous, then there exists a fixed point of $f$ in $Y$.
\end{lemma}

\begin{lemma}
\label{lem_3}
(Brouwer fixed-point theorem) For any continuous function of mapping a compact convex set to itself, there is a point $x_0$ such that $f(x_0)=x_0$.
\end{lemma}

\subsection{The Proof}

Define the set of players as $N=\{1,\ldots,n\}$, the set of time periods as $\mathcal{H}=\{1,\ldots,T\}$, and the set of reasoning levels as $\mathcal{L}=\{1,\ldots,L\}$. Let $A_{ilt}=\{a_{ilt1},...,a_{iltK}\}$ be the set of strategies for a player $i \in N$ with reasoning level $l \in \mathcal{L}$, and in time period $t \in \mathcal{H}$. Define the joint strategy as: $A = \prod_{i\in N, l\in \mathcal{L}, t\in \mathcal{H}} A_{ilt}$.

Denote function $\boldsymbol \pi:  \mathcal{L} \times \mathcal{H} \rightarrow \Delta_{\pi}$ as strategy. Assume players can choose positive integers from 1 to $K$. Thus, it follows that varying number of choice are available in one period. 
\begin{align*}
  \Delta_{K} = \left\{p_a = (p_{a_1},...,p_{a_{K}}) \in \mathcal{R}^{K}_{+} :p_a \geq 0, \;p_{a_1}+,...,+p_{a_{K}}=1\right\}.
\end{align*}
Denote $\Pi$ as the set of feasible policies, and $\pi \in \Pi$. By representing the QCH-IPL approach in the following perspective: 
\begin{align}
    (\pi_0, \ldots, \pi_m) = F\Bigl( G\Bigl(\bigl( CLR\left( \pi_0, \ldots, \pi_m, tr_i  \right) \bigr)_{i \in N}\Bigr) \Bigr), \label{eqn:fixed_pt2}
\end{align}
 we  can prove the existence of fixed points by proving there exists in $\Pi$ a fixed point of the combined strategy/reasoning level iteration $F(G(\cdot,tr_N))$, where $tr_N$ includes the behavior traces of all individuals. Since QBR is an equilibrium, and also a fixed point by Lemma \ref{lem_3}

Motivated by \citet{kaufman1998user}, our existence proof is based on Lemma \ref{lem_1}. The set of strategies can be represented as
\begin{align*}
  \Pi =\prod_{l\in \mathcal{L}, t\in \mathcal{H}, a\in \mathcal{A}} \Delta_{\pi},
\end{align*}
which can be viewed as a subset of $\Omega = \prod_{l \in \mathcal{L}, t \in \mathcal{H}, a \in \mathcal{A}} \mathscr{R}^2_{+}.$ To establish the continuity and compactness properties used to ensure the existence of a fixed point, we give $\Omega$ a topology. Here we choose Cartesian product topology applied to $\Delta_{\pi}$, which is given the usual topology of $\mathscr{R}^2_{+}$ \citep{munkres1975prentice}. It can be verified that $\Omega$ is a locally convex linear topological space from definitions in \citet{dunford1988linear}. Therefore, $\Pi$ is a convex subset.  $\Delta_{\pi}$ is closed and bounded, hence compact in $\mathscr{R}^2_{+}$. By Tychonoff theorem \citep{munkres1975prentice}, an arbitrary product of compact spaces is also compact in the Cartesian product topology, thus we show that $\Pi$ is both compact and convex. 

It remains to show that $F(G(\cdot,tr_N))$ is continuous. In the product topology \citep{munkres1975prentice}, to establish the continuity property, we need to show that each component $F_{lta}(G(\cdot,tr_N)):\Pi \rightarrow \Delta_{\pi}$ is continuous. $F$ is the strategy generation function, $F_{lta}$ is obtained by applying the quantal best response function $QBR^i(\lambda,\hat{\Delta}_l,\sigma^{-i}_0,\sigma^{-i}_1...\sigma^{-i}_{(l-1)})$,  where $\lambda$ is the quantal response parameter, $\hat{\Delta}_l$ is the normalized proportional of other agents for level $l$ agents. $\sigma^{-i}$ is the opponent's strategy, which is determined by the reasoning level distribution $d=G(\pi_0, \ldots, \pi_m, tr_N)$ and $(\pi_0, \ldots, \pi_m) = QBR(\pi)$. As $G$ and $QBR$ are continuous, $\sigma^{-i} = \sigma^{-i}(G(QBR(\pi), tr_N))$ is then continuous in $\pi$. By definition, the quantal best response function $QBR^i$ is continuous with respect to $\hat{\Delta}_l$ and $\sigma^{-i}$, and hence respect to $\pi$. Thus each component function $
F_{lta}$, and hence $F$ itself, is continuous on $\Pi$. According to Lemma \ref{lem_1}, a fixed point exists for \eqref{eqn:fixed_pt2}. \qed

\section{Contraction Proof}\label{B}
 
Motivated by \citet{kaufman1998user}, to establish property of contraction, we impose three additional restrictions on the level policies. The first restriction is, we only consider level policies in $\Pi$ for which each action vector varies continuously over time. Secondly, the set of level policies that are equicontinuous \citep{dunford1988linear}. In other words, each action vector of each policy in the policy space is Lipschitz continuous with respect to time, with a uniform Lipschitz constant characterizing all action vectors at all times. Ensuing an equicontinuous policy space allows us to use Euclidean norm for a particular policy and supremum norm for metric topology. The set of strategies in $\Pi$ whose component functions are Lipschitz continuous with a non-negative uniform Lipschitz constant are equicontinuous. The third restriction is to ensure the strategy set is closed in the iteration. However, the closure restriction can be complicated, since we should consider not only the functions, but also their interdependence. With sufficient closure restrictions, the existence of fixed points can be verified under time-continuity with more convenient topological conditions.
 
Denote the restricted strategy set as $\Pi^*$, by applying Lemma \ref{lem_2}, we can prove that a fixed point exists in $\Pi^*$ of the combined strategy/reasoning level iteration $F(G(\cdot,tr_N))$. \qed

\section{Inferred Reasoning Levels of All Participants}\label{C}
In tables below, we summarize the mean reasoning levels from all participants over all weeks.

\begin{table*}[htb!]
\caption{Average reasoning levels for participants 1-76 over all weeks.}
\label{tab:player_level1}
\centering
\setlength{\tabcolsep}{1.5pt}
\begin{tabular}{lccccccc|lccccccc}
\hline
\# $\setminus$ Wk & (1) & (2) & (3) & (4) & (5) & (6) & (7) & \# $\setminus$ Wk & (1) & (2) & (3) & (4) & (5) & (6) & (7) \\
\hline
1&1.74&7.60&11.69&18.00&13.91&8.00&6.32&39&2.37&10.15&7.49&12.86&11.32&9.29&6.90\\
2&2.50&9.63&10.56&5.40&16.91&8.33&9.02&40&3.14&8.78&17.89&11.43&10.65&7.75&7.88\\
3&1.21&8.68&12.36&9.67&9.78&14.61&4.71&41&1.23&12.13&14.38&12.12&11.53&11.03&10.50\\
4&1.20&12.63&14.83&3.01&7.81&10.00&7.00&42&1.21&14.41&14.93&12.29&5.15&8.00&6.68\\
5&1.71&13.00&15.00&6.85&12.09&8.00&10.53&43&3.38&11.62&15.41&12.09&4.20&8.95&7.00\\
6&1.53&8.23&18.00&15.70&18.00&6.12&7.91&44&1.17&5.44&10.34&14.26&7.57&8.28&5.02\\
7&1.90&8.52&17.26&2.93&10.87&10.77&5.63&45&2.38&10.49&13.74&11.21&10.97&8.41&9.97\\
8&2.89&14.50&14.05&6.65&17.45&13.39&8.52&46&1.33&11.23&19.30&10.14&18.00&7.00&10.62\\
9&1.53&9.03&16.04&8.99&10.39&7.08&5.68&47&1.09&12.66&12.11&10.44&10.68&10.00&4.02\\
10&1.00&9.44&14.93&10.59&13.26&6.98&6.49&48&1.57&10.27&17.24&11.88&7.90&10.47&10.31\\
11&2.57&9.92&12.91&9.85&14.00&1.74&1.80&49&2.84&10.07&9.98&12.09&7.83&8.01&11.00\\
12&1.26&13.31&15.47&12.09&14.47&7.00&9.44&50&3.35&11.80&14.98&12.09&7.58&11.20&8.39\\
13&1.00&8.27&11.43&7.42&3.89&11.42&6.85&51&3.81&13.80&15.05&5.68&2.01&8.47&4.69\\
14&2.51&7.38&7.58&8.48&1.74&4.66&1.80&52&3.00&14.00&18.00&12.00&16.16&8.00&11.00\\
15&2.11&6.72&19.19&15.10&16.13&6.12&11.84&53&3.12&14.02&18.70&14.00&15.00&6.79&7.29\\
16&1.52&15.48&19.31&30.35&15.73&17.67&12.71&54&1.84&7.00&19.17&10.09&8.90&9.93&8.39\\
17&1.52&14.55&18.90&16.88&7.21&13.31&7.88&55&1.78&14.65&14.00&14.47&13.06&4.32&7.87\\
18&2.39&12.67&12.94&9.38&8.14&8.08&9.39&56&2.44&7.13&14.93&7.79&1.00&2.76&3.84\\
19&1.15&10.58&16.61&11.36&12.68&6.71&5.47&57&1.94&10.43&15.00&7.80&15.09&8.39&6.00\\
20&2.42&7.32&19.05&16.61&14.14&10.57&8.41&58&3.00&13.47&16.23&10.67&14.39&11.75&6.67\\
21&1.00&6.21&16.20&24.00&9.78&6.70&4.65&59&2.73&11.45&13.75&11.06&11.52&6.25&9.68\\
22&1.00&11.25&16.13&12.45&16.68&7.23&8.22&60&3.66&13.42&15.84&5.68&15.61&9.85&9.06\\
23&3.02&13.61&12.59&10.52&7.01&5.64&6.40&61&2.07&9.46&12.49&7.21&14.76&12.36&7.68\\
24&3.37&7.79&11.77&13.82&7.01&8.20&9.05&62&3.75&9.33&14.70&12.61&18.19&7.09&11.17\\
25&1.74&12.00&14.93&12.09&14.00&6.48&5.42&63&3.68&14.47&19.80&18.42&7.00&8.36&8.39\\
26&1.38&10.61&5.01&7.17&9.41&7.51&6.20&64&1.83&12.13&12.99&11.72&10.35&12.39&5.45\\
27&3.00&9.70&14.16&12.92&12.61&10.32&8.51&65&2.34&7.76&18.69&12.58&11.96&10.30&6.22\\
28&3.97&12.30&14.88&10.00&6.00&10.95&8.85&66&3.75&4.93&17.56&8.49&14.00&4.52&9.21\\
29&0.65&9.64&20.00&10.39&5.29&3.18&6.00&67&1.73&12.66&17.83&13.44&11.65&13.55&12.04\\
30&1.64&8.51&11.88&16.57&8.57&8.11&10.67&68&3.54&9.71&16.32&15.09&8.83&10.04&6.96\\
31&2.23&9.13&10.17&13.28&11.20&8.61&5.67&69&2.34&8.75&18.94&28.10&16.93&10.61&6.00\\
32&1.41&7.00&17.41&10.04&11.74&3.37&5.46&70&4.00&7.82&15.73&10.48&12.93&10.41&9.05\\
33&3.49&4.82&8.54&4.67&5.17&8.29&5.32&71&4.00&14.57&18.05&13.28&8.05&6.37&5.74\\
34&1.07&9.99&8.84&1.00&6.65&7.12&8.59&72&4.00&14.40&16.43&17.91&6.37&10.61&6.98\\
35&2.39&11.49&14.93&13.22&12.19&10.94&7.26&73&2.67&8.54&17.14&12.27&6.57&7.45&8.06\\
36&1.87&9.88&1.00&1.00&1.00&1.00&2.62&74&4.00&9.00&15.08&14.66&13.97&7.78&5.76\\
37&4.36&7.21&18.27&10.93&15.02&8.00&11.00&75&0.91&1.84&18.04&14.86&15.14&11.11&7.56\\
38&2.88&17.00&14.93&9.59&14.00&9.22&9.61&76&4.00&8.09&18.33&14.85&10.23&7.07&11.16\\
\hline
\end{tabular}
\end{table*}

\begin{table*}[htb!]
\caption{Average reasoning levels for participants 77-152 over all weeks.}
\label{tab:player_level2}
\centering
\setlength{\tabcolsep}{1.5pt}
\begin{tabular}{lccccccc|lccccccc}
\hline
\# $\setminus$ Wk & (1) & (2) & (3) & (4) & (5) & (6) & (7) & \# $\setminus$ Wk & (1) & (2) & (3) & (4) & (5) & (6) & (7) \\
\hline
77&4.00&12.13&12.93&1.00&11.02&5.01&6.00&115&2.00&10.59&24.18&25.65&12.17&8.34&11.69\\
78&3.42&4.62&4.71&8.25&7.78&7.77&3.63&116&2.28&9.64&19.69&11.66&4.66&10.90&6.93\\
79&2.00&2.80&8.78&8.70&14.56&10.73&3.52&117&1.60&7.15&13.47&10.43&13.05&9.10&7.71\\
80&3.60&10.24&17.97&7.62&8.77&2.02&8.33&118&2.39&2.58&17.23&4.46&10.83&11.64&5.16\\
81&2.61&3.90&1.00&5.60&8.16&3.79&3.69&119&2.20&5.61&15.90&13.32&6.87&10.00&5.08\\
82&3.13&11.17&16.40&9.57&8.77&9.45&9.35&120&2.00&5.89&11.76&2.42&7.69&10.06&10.62\\
83&4.00&10.73&1.00&11.54&15.96&10.74&8.44&121&2.34&14.67&15.44&10.13&5.42&7.14&5.63\\
84&2.81&6.42&15.79&11.93&9.07&8.50&10.78&122&2.90&6.20&18.99&8.76&12.48&5.05&8.29\\
85&2.99&7.57&16.07&2.50&6.65&7.21&5.55&123&3.16&12.25&15.81&18.00&6.00&11.00&8.39\\
86&3.93&10.05&13.29&11.89&12.04&8.50&5.52&124&2.88&12.00&16.34&10.19&4.90&8.83&9.64\\
87&2.14&9.09&3.98&5.41&3.67&12.53&3.79&125&2.21&9.70&16.03&14.35&11.25&10.98&5.88\\
88&1.87&5.75&14.56&4.78&8.65&7.24&10.54&126&0.26&0.00&0.14&0.00&0.00&0.00&0.00\\
89&2.00&10.10&16.77&6.96&2.79&5.01&8.01&127&3.35&7.66&16.60&11.89&7.22&11.00&9.00\\
90&2.00&9.72&15.08&8.89&12.94&9.39&7.97&128&1.00&1.00&1.00&1.00&1.00&1.00&1.70\\
91&2.00&7.04&18.00&12.00&3.35&9.17&8.00&129&1.05&11.49&16.04&8.96&14.31&12.79&5.38\\
92&2.79&8.79&14.11&10.08&10.25&9.17&9.61&130&2.00&10.00&19.64&12.95&18.00&7.00&10.62\\
93&1.26&14.00&18.00&12.36&16.16&8.00&8.41&131&1.64&13.88&15.34&6.52&14.57&11.20&9.56\\
94&2.00&12.37&17.26&12.31&15.00&12.00&8.00&132&0.55&11.87&16.47&12.00&16.16&7.00&10.24\\
95&2.00&9.59&17.94&13.83&13.22&5.69&10.62&133&2.68&6.42&17.82&15.98&10.41&9.64&8.32\\
96&1.63&13.08&18.86&13.19&15.42&9.10&7.12&134&3.39&10.74&18.19&9.55&11.28&7.90&10.82\\
97&2.00&11.71&16.90&13.64&16.63&11.10&1.00&135&3.27&9.39&16.59&11.66&14.48&7.15&6.12\\
98&2.00&7.13&9.61&10.59&15.29&5.49&6.78&136&1.33&8.50&14.99&6.30&15.98&11.64&8.52\\
99&2.00&7.99&15.89&10.59&9.78&8.76&6.36&137&2.36&8.82&14.93&16.11&12.52&8.14&5.59\\
100&2.00&8.22&10.51&14.86&9.20&7.17&10.12&138&2.32&7.71&11.66&11.07&14.54&7.62&9.51\\
101&2.00&14.05&14.59&10.79&9.48&11.00&8.39&139&1.58&8.00&17.40&3.67&5.44&5.37&1.41\\
102&2.00&9.01&15.41&10.67&4.54&8.75&4.71&140&1.90&11.29&14.56&33.01&12.64&14.57&11.04\\
103&1.63&14.55&19.14&14.08&6.00&10.83&10.20&141&1.77&12.66&18.04&8.52&1.00&11.18&13.45\\
104&1.85&11.10&17.75&12.74&13.46&10.01&9.54&142&1.90&7.88&17.61&12.26&18.00&10.09&10.41\\
105&2.00&14.88&18.49&9.79&13.59&8.84&8.58&143&3.17&7.64&12.83&5.12&14.83&13.22&10.67\\
106&2.72&8.32&17.20&10.19&13.19&11.25&7.93&144&2.33&15.39&15.48&9.16&11.57&13.38&9.22\\
107&2.73&15.96&19.31&10.34&10.95&10.10&8.44&145&4.00&14.04&18.64&8.51&16.21&8.77&7.37\\
108&2.00&7.15&9.48&12.88&18.00&10.40&8.52&146&2.21&12.46&24.28&17.28&12.57&13.47&9.44\\
109&1.10&9.64&17.70&4.42&6.37&6.16&3.90&147&4.74&6.14&18.87&12.49&13.86&8.70&8.42\\
110&2.00&11.10&18.70&14.00&15.00&12.00&8.00&148&2.34&8.36&16.91&7.98&9.89&6.18&7.75\\
111&1.55&11.78&17.42&7.11&5.43&6.60&1.00&149&1.99&14.00&18.00&12.00&15.22&9.67&9.48\\
112&0.29&14.52&15.17&12.09&11.15&9.87&8.17&150&1.00&10.09&18.22&14.44&4.06&6.31&9.13\\
113&1.35&15.57&8.05&1.00&6.00&11.00&8.39&151&3.07&13.72&20.43&16.77&15.61&7.41&10.74\\
114&0.00&4.61&16.34&9.49&9.47&5.50&3.85&152&1.51&13.45&12.66&8.61&4.20&3.99&4.82\\
\hline
\end{tabular}
\end{table*}

\end{document}